\documentclass[11pt,twocolumn]{article}

\usepackage{times}

\usepackage[labelfont=bf]{caption}

\usepackage[top=0.5in,bottom=0.75in,left=0.5in,right=0.5in]{geometry}

\usepackage{url}

\usepackage{subfig}

\usepackage{graphicx}

\usepackage{amsmath}

\graphicspath{{figures/}{plots/}{.}}

\usepackage[numbers]{natbib}
\usepackage[bookmarksopen,bookmarksnumbered,colorlinks=true,allcolors=blue]{hyperref}

\usepackage{xcolor}

\usepackage[compact]{titlesec}

\usepackage{helvet}
\usepackage{sectsty}
\allsectionsfont{\sffamily}

\begin{document}


\title{\sffamily Comparing Overlapping Data Distributions Using Visualization}

\author{\sffamily Eric Newburger and Niklas Elmqvist\\ 
\scriptsize\sffamily University of Maryland, College Park, MD, USA}

\date{\sffamily April 2023}

\maketitle

\begin{abstract}
    We present results from a preregistered and crowdsourced user study where we asked members of the general population to determine whether two samples represented using different forms of data visualizations are drawn from the same or different populations.
    Such a task reduces to assessing whether the overlap between the two visualized samples is large enough to suggest similar or different origins.
    When using idealized normal curves fitted on the samples, it is essentially a graphical formulation of the classic Student's t-test.
    However, we speculate that using more sophisticated visual representations, such as bar histograms, Wilkinson dot plots, strip plots, or Tukey boxplots will both allow people to be more accurate at this task as well as better understand its meaning.
    In other words, the purpose of our study is to explore which visualization best scaffolds novices in making graphical inferences about data.
    However, our results indicate that the more abstracted idealized bell curve representation of the task yields more accuracy.
\end{abstract}

\textbf{Keywords:} Graphical inference, visual statistics, statistics by eye, fitting distributions, crowdsourcing.

\begin{figure*}[tbh] 
    \centering
    \subfloat[Overlapping bell curves.]{
        \resizebox{0.45\textwidth}{!}{\includegraphics{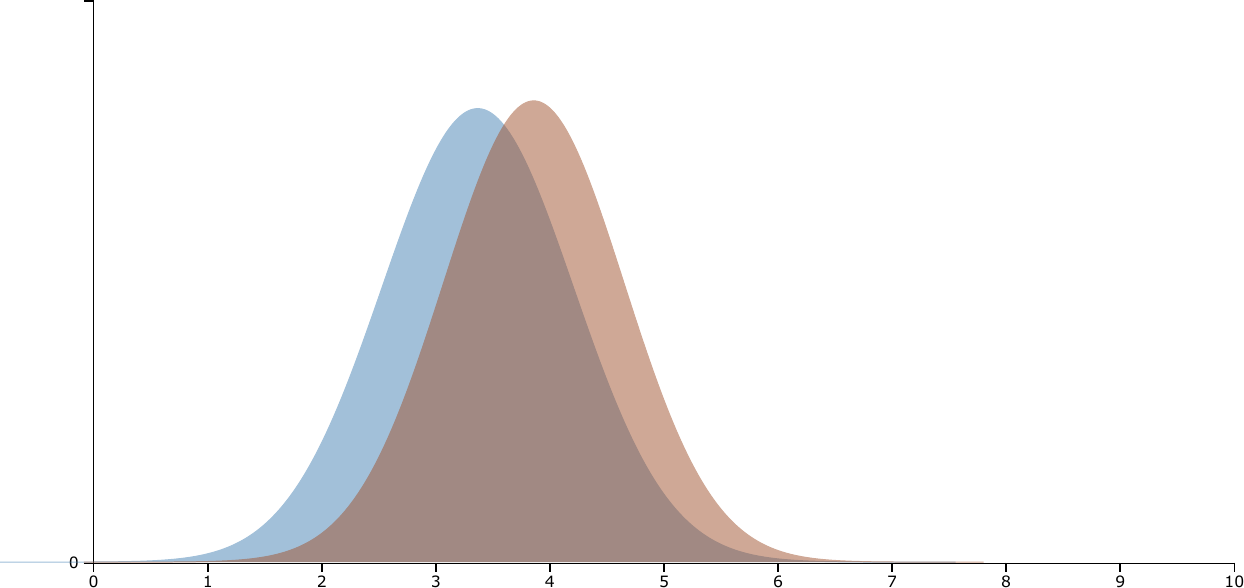}}
        \label{fig:stimulus-bell}
    } 
    \subfloat[Wilkinson dotplots.]{
        \resizebox{0.45\textwidth}{!}{\includegraphics{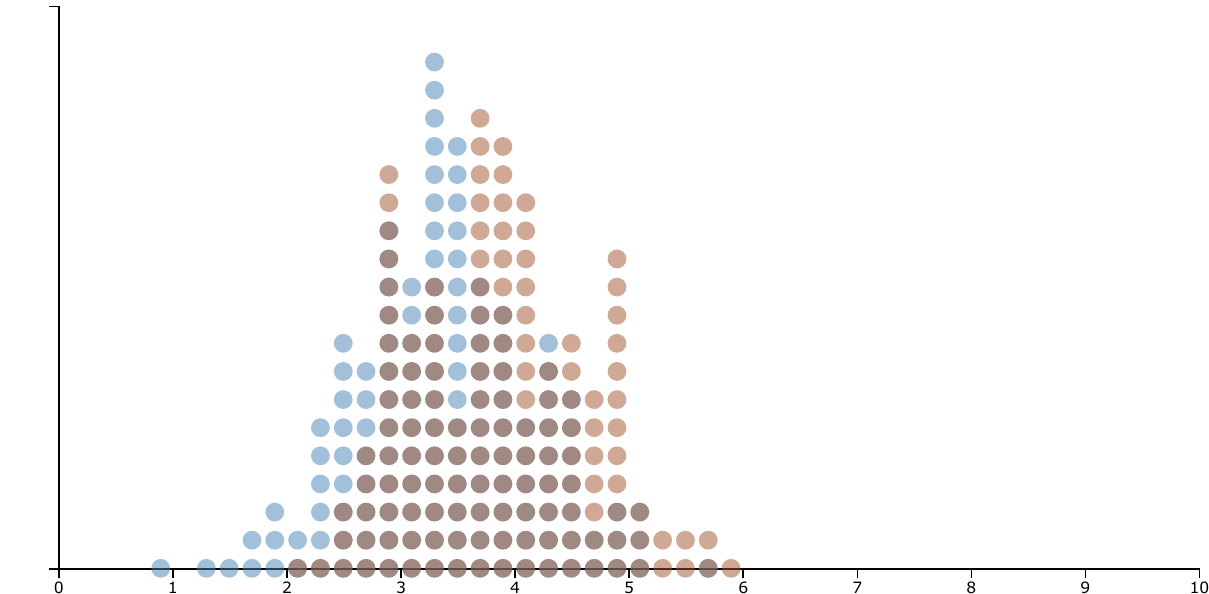}}
        \label{fig:stimulus-dot}
    }
    \\
    \subfloat[Bar histograms.]{
        \resizebox{0.45\textwidth}{!}{\includegraphics{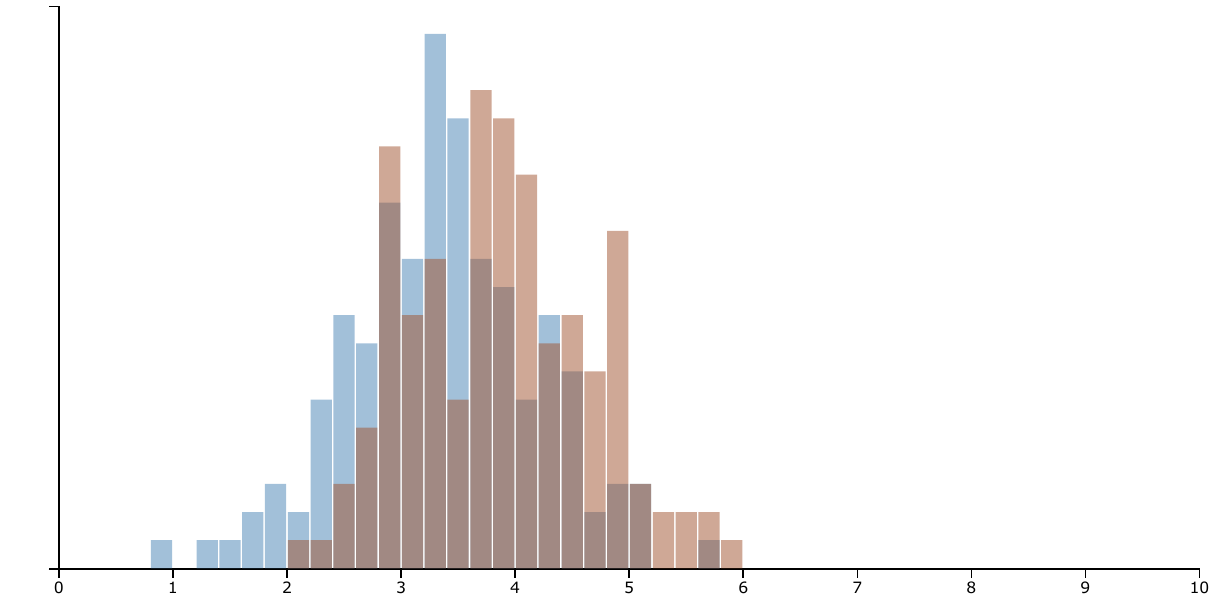}}
        \label{fig:stimulus-histogram}
    }
    \subfloat[Stacked bar histograms.]{
        \resizebox{0.45\textwidth}{!}{\includegraphics{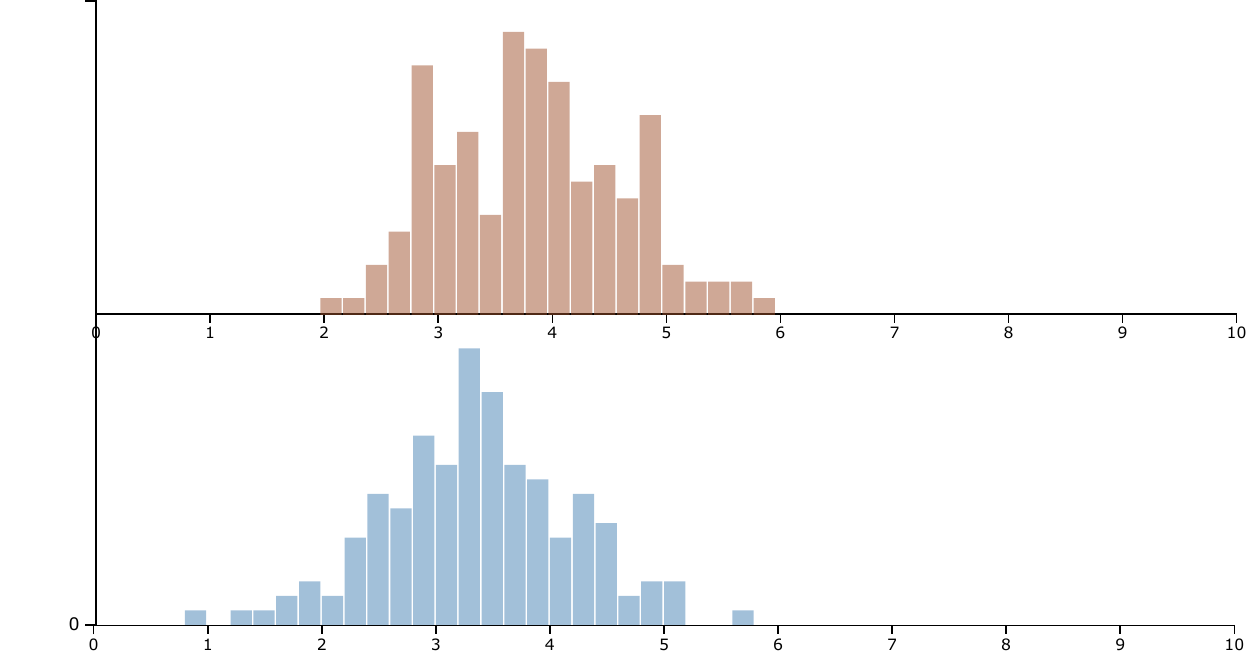}}
        \label{fig:stimulus-stacked}
    }
    \\
    \subfloat[Box plots.]{
        \resizebox{0.45\textwidth}{!}{\includegraphics{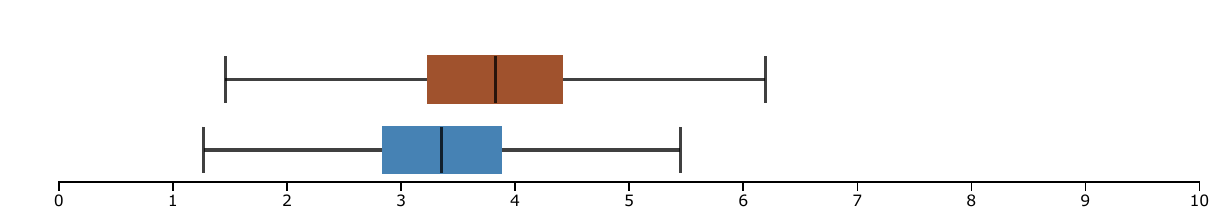}}
        \label{fig:stimulus-box}
    }
    \subfloat[Strip plots.]{
        \resizebox{0.45\textwidth}{!}{\includegraphics{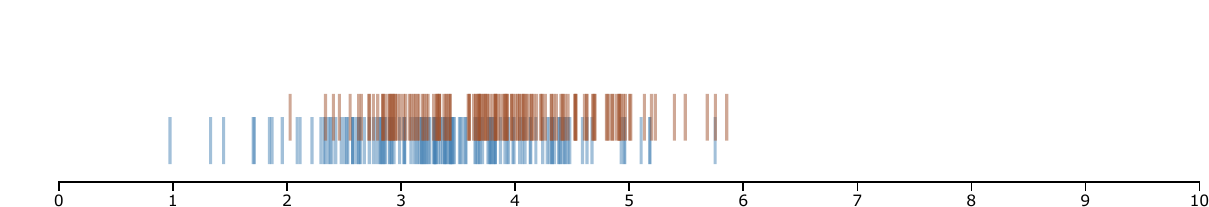}}
        \label{fig:stimulus-strip}
    }
    \caption{\textbf{Visualizations of data distributions.}
    Two data samples ($n=144$ items each) visualized in the same chart, allowing for participants to determine whether or not the samples were drawn from the same or different populations.
    The samples were visualized using six different visual representations.
    }
    \label{fig:teaser}
\end{figure*}
    

\section{Introduction}

Central to \textit{confirmatory data analysis} is the concept of determining whether two data samples are drawn from the same or different populations.
For the simplest of such inferential statistical tests---the Student t-test~\cite{Student1908}---this essentially amounts to fitting a normal distribution to each sample and then determining the overlap between them, adjusting expectations for the amount of overlap in light of the size of the two samples. 
However, even t-tests are not understandable to the general population, and this is doubly true for more sophisticated statistical tests.
Despite this, all of us routinely have to do analogous tasks in our daily lives, such as when determining whether a specific credit card bill is out of the ordinary---potentially indicating credit card fraud---or when assessing and comparing a child's, employee's, or public figure's performance in school, at work, or in the public eye.

In this paper, we study how graphical formulations of t-tests can support novices with no specialized statistical training in assessing the differences between two or more data samples. 
Assuming normally distributed data, a straightforward way to achieve this is to fit normal distributions to each data sample (of known sample size), and then visualize the samples as overlapping bell curves.
A user can then manually determine whether they think the two curves represent the same or different underlying populations.
Such a graphical formulation corresponds to the classic t-test.~\cite{Student1908} Such formulation is similar to looking at overlapping confidence intervals, but with additional visual information from the specific shape of each curve.
However, a t-test can be a relatively blunt instrument, so we are also interested in how more detailed (e.g., less abstract) data visualizations might better support both user performance as well as understanding of these statistical tests.
For this reason, we broaden our investigation to also include the typical visual representations suggested in the literature for visualizing data distributions,~\cite{journals/tvcg/CorrellLKS19} including bar histograms (overlapped and stacked), Wilkinson dot plots, strip plots, and Tukey boxplots. 

We conducted a crowdsourced study to understand the above question: can people with no statistical training use these graphical formulations to perform a t-test? 
In other words, can they determine whether or not two samples are drawn from the same or different populations?
A total of $N = 212$ participants were given a sequence of trials where they were asked this question under varying conditions: different visualizations (overlapping bell curves, stacked bell curves, bar histograms, dot plots, strip plots, and boxplots) and data sizes (36, 144, or 1,000 items per dataset). 
A single trial typically lasted less than 10 seconds, allowing us to collect a large number of trials per participant.
In the interest of transparency in research, the study was pregistered on OSF: \url{https://osf.io/b5dyf}.

Our results indicate, not surprisingly, that this is a difficult task, and that people regularly overestimate how much divergence is needed to indicate that two samples were likely drawn from different populations.
Furthermore, the task gets particularly difficult---approaching the random 50\% accuracy---when the difference between samples is small, and easier for very different samples.
What is more surprising, however, is that more detailed visualizations that limit aggregation yield less accuracy than the highly aggregated boxplot and idealized bell curve representations, particularly for higher data sample sizes.
It appears as if large samples obscure rather than improve judgments, likely due to overplotting.
Highly aggregated representations yield an opposite effect---large samples result in better accuracy---which supports this interpretation.
In other words, our results show that graphical inferences of t-tests can be effective in some circumstances, but that abstracting and aggregating the data generally yields better results.

\section{Background}

John Tukey helped launch the modern era of data visualization with his publication of \textit{Exploratory Data Analysis} in 1977.~\cite{Tukey1977} 
In it, he argued that data visualization provided researchers powerful means for detecting outliers or errors, identifying a multitude of patterns, and recognizing structures within their data, all of which made it an ideal analytic tool for hypothesis formation.
Tukey went on to recommend the use of traditional, equation-based statistical methods for testing and confirming hypotheses.
He believed that different methods should be used for hypothesis formation and confirmation, to ensure the analyst avoids the trap of circular reasoning.
Given the flexibility visualization affords researchers in finding the unexpected within data, he argued that visualization tools are particularly well suited for an \textit{exploratory} role, where the data is not known a priori and questions and hypotheses are generated over time.
This is a stark contrast to the \textit{confirmatory} role typically played by classic statistics, where such questions and hypotheses are answered through more or less rigorous testing.

\subsection{Traditional Confirmatory Analysis}

Since Tukey---one of the premier statisticians of the 20th Century---endorsed this use of data visualization, it has gained growing acceptance, not just for communicating findings, but also as an exploratory method in its own right.
However, while visualization may be seen as useful by the research community, traditional confirmatory statistics is generally seen as essential.
Tests of statistical significance are a cornerstone of statistical and scientific practice.
While still including elements of subjective, arbitrary assumptions, such as the choice of a significance level (alpha value), they nonetheless provide standards for decision making that push interpretations of experimental results closer to objectivity.
Ideally, they act as a check when our enthusiasm for a favored hypothesis might otherwise have motivated us to make claims based upon results that could just as easily have been explained by random noise in the data.
This indispensable discipline comes in myriad forms, such as t-tests and ANOVAs, where each is suited to different analytic methods but all are founded upon a five-step process:

\begin{enumerate}
\item The researcher forms an \textit{expected value} based upon their null hypothesis;
\item The researcher selects a \textit{level of improbability} that they consider statistically significant (often $\alpha = 5\%$);
\item The researcher \textit{extracts an actual value from the data} and compares it to the value expected under the null hypothesis, thus creating a test statistic;
\item The researcher \textit{applies appropriate probability calculations} (which account for assumptions tailored to the particular data collection in question) to determine how unlikely it was that random noise would result in a test statistic as large as the one they found; and 
\item If the result is so unlikely to have happened by chance that it breaches the pre-chosen significance limit, the researcher \textit{rejects the null hypothesis} and declares that they have support for their favored hypothesis.
\end{enumerate}

This process affords the researcher a sense that their results are more than mere anecdote, but instead, form  scientifically rigorous evidence.
Yet, applying the process correctly can be a challenge.
Equation-based statistics include subjective elements at nearly every stage, from the selection of an initial hypothesis and its mathematical expression, to the choice of a p-value limit, to the creation of experimental design, to the choice of the right probability assumptions to form a test statistic.
These elements make it possible for people to, for example, ``p-hack''---manipulating data and data analysis to find statistically significant effects, a highly questionable analysis practice---either knowingly or by mistake.  
Ironically, the wide availability of automated computation tools, such as spreadsheets and other statistical software packages, may compound the problem.
These tools turn statistical procedures into magic black boxes, which consume data and spit out results without requiring users to understand all the assumptions which undergird those results.
Moreover, it becomes simple to churn out multiple analytic variants on a dataset, until one pops out that looks promising.
For example, when conducting 100 experiments, we should expect through random chance to find five results that are significant at the 5\% level.
Reporting these hits without the misses gives a false impression of evidence supporting a finding when there is none.
This kind of error---which can be entirely unintentional---may be at the root of the \textit{replicability crisis}, which has plagued many scientific fields, in particular medicine and psychology, but also visualization~\cite{Kosara2018} and HCI.~\cite{DBLP:conf/chi/EchtlerH18}
Researchers have attempted to address this crisis through pre-registration, registration reports, replications of past work, etc. 

Despite these issues, the application of equation-based statistical methods to hypothesis confirmation has become a linchpin in advancing our understanding of the world.
Tables of results with p-values are ubiquitous within published scientific work.
This elevates equation-based statistics in the scientific community, while leaving visual analyses in a secondary role.
For example, during their undergraduate studies, most science majors will take an equation-based statistics course by requirement, but outside of a few select fields (such as data science), they are unlikely to learn visualization in any formal way.

As data visualization practitioners, we might address this situation in two ways.
First, we can attempt to demonstrate that visual analyses can act in the hypothesis confirmation role.
Second, we can document affordances provided to researchers through visual confirmation analyses which provide some advantage separate from those provided by equation-based methods.  

\subsection{Visual Confirmatory Analysis}

Tukey's reasoning on using one set of methods for hypothesis formation and a second set for hypothesis confirmation, comes with no requirement that visualization fill the first role.
Should we develop sufficiently flexible equation-based methods for hypothesis generation, we could conceivably pair them with visual analyses for hypothesis confirmation.
Or we might create two different kinds of visual analyses, one set for exploration, the other for hypothesis testing.
But what would a data visualization designed for hypothesis confirmation look like?  

Just as we can describe statistical tests in five steps, we can describe what goes into those as four intellectual products:

\begin{enumerate}
    \item\textbf{Precise measures of some phenomena} (precise relative to the effect we want to explore), including an observed value of interest.
    These measures emerge from our experimental designs, and comprise the data.

    \item\textbf{Distributions of idealized data populations} from which the data could have been drawn given the observed data distributions and sample scheme.
    These reflect our assumptions about the true population.
    Classical parametric tests are defined for specific distributions; often a Gaussian (normal) distribution, in which case we use our observed data to calculate a mean and standard deviation.
    We treat these derived values as estimates of the unseen, true values for the total population, that is, we assume they describe the central tendency and spread of that larger population.

    \item\textbf{An expected value.}
    This derives from the null hypothesis, based usually upon an assumption of random behavior as opposed to behavior driven by some combination of underlying forces we hope to detect.
    We typically calculate this by considering what would happen should we run our experiment in an environment in which the effect size for the forces we wish to explore were set at zero.

    \item\textbf{A table of probabilities} expressing how unlikely we are to measure results that deviate by this or that amount from the expected value, given the sample size.
    This table derives from a function built with the same assumptions that underlie the idealized data distributions, paired with a model to account for our sampling processes.
\end{enumerate}

Each of these products are fundamentally quantitative, and thus, could be expressed with data visualization.
Therefore, at a minimum, we might create visualizations of hypothesis tests by performing a one-to-one mapping of the math onto shapes and lines, just as we can, for example, represent regression equations with straight lines drawn through a data field on a scatterplot.
But it still remains to be seen whether we, in fact, perform a regression with our eyes, or whether we are instead performing some visual proxy for the regression or other mathematical operations.
If mere proxies, then being able to perform a literal translation of quantities into geometries tells us little about whether we might usefully decode such images with our eyes, nor what affordances such an approach might provide to the prospective researcher.
That is what this research will attempt to explore for graphical formulations of a classic statistical test.

\section{Related Work}

We review prior evidence on graphical inference below, discuss methods for evaluating these phenomena, and discuss visualization techniques for data distributions.

\subsection{Graphical Inference in Statistics}

Statistics and visualization being different disciplines is a relatively new development. 
Virtually all statistical workflows involve creating graphical representations of data.~\cite{Cleveland1993}
In fact, for some, such as exploratory data analysis,~\cite{Tukey1977} the practice is central.
It thus follows that \textit{graphical inference} using such representations is equally common as making inferences from algorithmic representations. 
For example, rather than formal testing normality, many practicing statisticians will instead ``eyeball'' the sample in a quantile-quantile plot (QQ-plot) against the normal distribution to ensure that the samples fall on a line (indicating high correlation with normality in the structure of the sample).

Graphical inference is also not a new practice.
An early example is Scott et al.'s seminal work from 1954,~\cite{Scott1954} where statisticians validated astronomical models asking people to compare artificial star charts to real charts.

\subsection{Graphical Inference in Visualization}

Compared to dedicated statistical tests, a strong appeal of a visual representation is that it can support more exploratory data analysis without the need for preconceived questions or prior knowledge about the dataset~\cite{Tukey1977}; e.g., to support \textit{hypothesis generation} rather than confirmatory analysis.

The visualization community has only recently begun to study formal graphical inference.
Correll et al.~\cite{Correll2012} investigated how line graphs can be designed to enable detecting maximum averages in time-series data.
Then Albers et al.~\cite{Albers2014} generalized this idea to six aggregate tasks for eight different time-series visualizations.
Aigner et al.~\cite{journals/cgf/AignerRH12} augmented line graphs with color to better support visual statistics.
Finally, Fuchs et al.~\cite{conf/chi/FuchsFMBI13} designed line glyphs to support higher-level aggregate tasks.
Most recently, Correll and Heer~\cite{conf/chi/CorrellH17} studied people's ability to fit trend lines to bivariate visualizations in a crowdsourced experiment, in essence performing regression analysis.
All of these studies were inspirational for our work even if our approach and experiment are different.

Other applications of graphical inference in visualization involve additional forms of data.
For multi-class scatterplots, Gleicher et al.~\cite{Gleicher2013} found that mean value judgments are reliably accurate independent of the number of points and conflicting encodings.
A related study asked participants to compare the average height of two groups of bar charts and found that accuracy was improved by increasing the number of bars but declined as variance increased.~\cite{Fouriezos2008}
Cumming and Finch recommended the use of overlapping confidence intervals (drawn as error bars) as a direct path to visual inference, and recommended specific design features for constructing displays with them.~\cite{Cumming2005}
However, Belia et al.~\cite{Belia2005} found that researchers commonly have ``severe misconceptions'' about the meaning and proper interpretation of confidence intervals.  
Correll and Gleicher~\cite{Correll2014} conduct crowdsourced experiments that yielded redesigns of error bars in bar charts, showing how violin or gradient plots produce insights more aligned with statistical inference.
Finally, Henkin and Turkay~\cite{DBLP:journals/tvcg/HenkinT22} study natural language verbalizations describing correlation in scatterplots, yielding a wide vocabulary with significant commonality for scatterplots exhibiting high correlations.

Uncertainty visualizations are commonly used for magnitude estimation or decisions through visual comparison.
Kale et al.~\cite{DBLP:journals/tvcg/KaleKH21} recently presented results on understanding satisficing strategies for eight different such uncertainty representations. 
These findings are highly relevant to our work, but our choice of visual representations only overlap for dotplots and bell curves, and we also connect our work directly to dichotomous testing using a Student t-test.

\subsection{Visualizing Distributions}

Anscombe's Quartet~\cite{Anscombe1973} shows us that descriptive statistics are often insufficient even for simple tasks, and also highlight the power of representing the raw data in visual form. 
For large-scale data distributions, displays which incorporate some degree of data aggregation can still allow the viewer to inspect the data for flaws, missing values, or noise. 
One common such aggregated representation for univariate distributions is the bar histogram, which counts data occurrences in discrete bins and visualizes them using bars.
Sizing these bins thus becomes a primary concern, and several rules of thumb how to define them exists.
We here employ Sturges's rule,~\cite{Scott2009} which is based on the assumption that the distribution to be binned is Gaussian.
This is appropriate for our experiment since all of our trial datasets are drawn from a normal distribution.

Of course, histograms have flaws, most of them related to bin size and bin number.~\cite{journals/tvcg/CorrellLKS19}
Even disregarding binning, the aggregating nature of histograms means that the bars do not convey any information about the absolute number of cases in the distribution.
This is where representations such as Wilkinson dotplots,~\cite{Moon2016} where each discrete item in a bin is represented as a circle stacked on top of other items in the bin, become potentially useful. Other alternate representations such as dotplots, strip plots, density plots, violin plots, and gradient plots~\cite{conf/chi/KayKHM16} vary in their degree of aggregation, but lack the easy familiarity of histograms.

\subsection{Graph Comprehension}

Finally, our work is also relevant to the general topic of \textit{graph comprehension},~\cite{Pinker1990, Shah2002} which studies how visual representations exploit cognitive and perceptual mechanisms in an effective manner.
The mere fact that graphs are a recent invention suggests that their effectiveness stems from clever use of preexisting such mechanisms.~\cite{Pinker1990}
For example, Shah and Freedman~\cite{Shah2011} showed that using different visual representations---bar or line graphs---of the same data impacts the kind of inferences can be drawn from the data.
More importantly, they also found that the overall effectiveness is strongly dependent on the \textit{graphicacy} (graphical literacy)~\cite{Pinker1990} of the individual. 
Such findings also indicate how charts and graphics can be both evaluated using metrics such as aesthetic value, learning efficiency and performance efficiency,~\cite{Barnes2016} as well as what cognitive abilities and expertise are involved in the design of such graphs.~\cite{Barnes2017}
Building on a body of such work, Franconeri et al.~\cite{Franconeri2021} present a comprehensive overview of research-backed guidelines for creating effective visualizations designed for communicating data to its viewers.

Of course, the same perceptual and cognitive mechanisms that make chart comprehension effective can also be deceived to yield false insight.
There is a significant body of work in the visualization and statistics literature on such deception---intentional or not---using data visualization and how it can be mitigated.
Cairo~\cite{Cairo2019} provides a high-level summary of such approaches.

\section{Study: Overlapping Bell Curves}

The fundamental research question tackled in this paper is how well people can determine whether two data samples are drawn from the same or different populations, and whether their ability depends upon the visual representation used.

To answer this question, we conducted a crowdsourced experiment where participants were asked to make a decision about whether two samples visualized using idealized bell curves, boxplots, overlapping bar histograms, stacked bar histograms, Wilkinson dotplots, or strip plots, represented a statistically significant difference (t-test at the $p = 0.05$ level).
Since statistical significance is heavily dependent on the size of each sample, we grouped trials by the sample size of test datasets, and included an instruction page about that specific sample size, followed by two examples.
Where possible, trials included visual representations to convey the scale of the underlying data, whether directly, as in the case of strip plots or Wilkinson dotplots, or as a labels on the vertical axis for histograms and bell curves.
Boxplots included no sample size indicators on specific trials.     
This procedure is similar to Cumming and Finch’s recommendation for comparing overlapping confidence intervals.~\cite{Cumming2005}
However, the visualizations we use incorporate, to varying degrees, more visual information, and thus are less abstract than mere error bars.  

We preregistered this study on OSF: \url{https://osf.io/b5dyf}.
Furthermore, our OSF repository contains complete supplemental material for our study: \url{https://osf.io/876ur/}.
Below we review our methods, followed by our results in the next section.

\subsection{Crowdsourcing Rationale}

We designed this study to engage members of the general population in graphical inference tasks. 
For this reason, we opted to conduct our study using Amazon Mechanical Turk (MTurk).
Unfortunately, the use of Mechanical Turk means that we have little control over participant demographics and expertise as well as their computer hardware.
However, prior work by Heer and Bostock~\cite{Heer2010} has shown that perceptual tasks often yield acceptable results from crowdsourcing.

We argue that our study, while higher level in scope than low-level graphical perception, nevertheless engages respondents in a quick perceptual task that takes less than 10 seconds per trial. 
Furthermore, crowdsourcing enables us to collect data from a large number of respondents representating a good cross-section of society.

\subsection{Participants}

We originally planned to recruit a total of 500 participants (Turkers), from whom we would solicit answers to a bare minimum of trials each.
However, our approach evolved as we chose Turk Master Workers to help ensure high quality responses; we employed fewer of these ``premium'' Turkers to answer more trials each.

In the end, we recruited a total of 212 responders, limiting participation of Turkers to those with a proven track record on the site, that is, workers with 50 or more prior tasks done for other employers on the site, and with a 95\% acceptance rate for their work.
We also limited participation to people from the United States due to tax and compensation restrictions imposed by our IRB.
We screened participants to have no color vision deficiency (self-reported).

Data collection for the six visualization types took place through six separate surveys, each identical in question order and instructions, varying only in the visualization type presented.
Individuals were prevented from participating in the same survey multiple times.
However, it was possible for an individual to complete surveys for more than one visualization type.
For the purpose of these analyses, we treat such cases as separate responses. 

All participants were ethically compensated at a rate chosen to be consistent with an hourly wage of at least \$15/hour (the U.S.\ federal minimum wage in 2020 was \$7.25).
More specifically, the payout was \$2.50 per session, and with a typical completion time of 545 seconds, this yielded an hourly wage of approximately \$16.50/hour.

\subsection{Apparatus}

The study was distributed remotely through the user's web browser.
This also meant that were not able to control the specific computer equipment that the participants used.
We required all devices to be personal computers (laptop or desktop); tablets or smartphones were disallowed due to the limited screen space available on such devices.

\begin{figure}
    \centering
    \includegraphics[width=\linewidth]{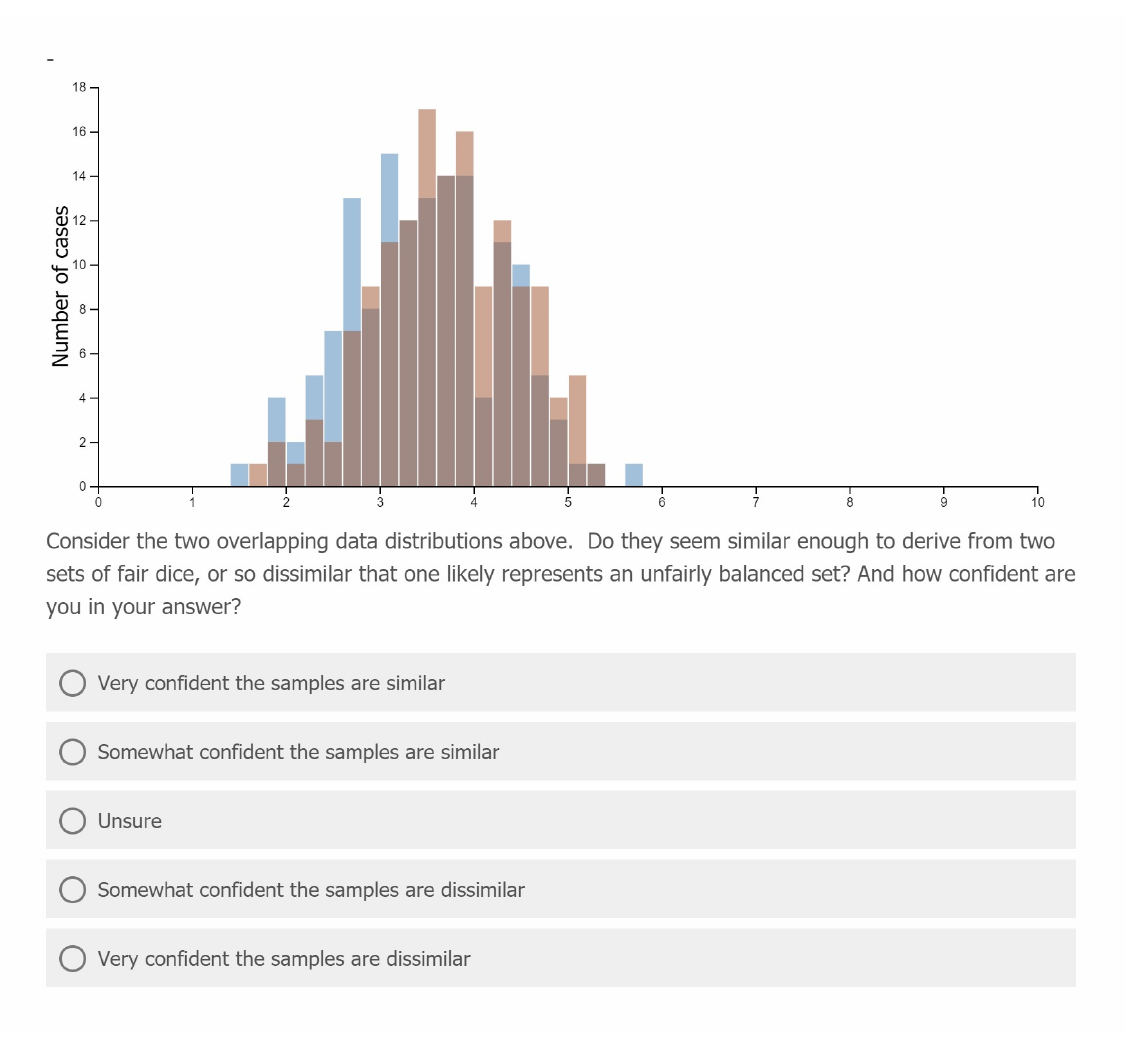}
    \caption{\textbf{Example of a trial.}
    In this example, the sample size $S$ is 144 items and the visual representation $V$ is bar histograms.
    The blue bars represent sample $A$ and the tan bars represent sample $B$; the brown color is their intersection.
    The vertical axis conveys the scale of each data sample.
    }
    \label{fig:trial}
\end{figure}

\subsection{Task}

Our study consisted of a sequence of trials involving a single task: determining whether two data samples visualized in a non-interactive chart in the user's browser represented the same or different source populations. 
The participants were given detailed instructions prior to beginning these trials which put the task in terms of comparing different sets of dice, one known to be fair and another uncertain.  In each trial, they were shown the chart as well as the following prompt (Figure~\ref{fig:trial}):

\begin{quote}
    \textit{Consider the two overlapping data distributions above. Do they seem similar enough to derive from two sets of fair dice, or so dissimilar that one likely represents an unfairly balanced set? And how confident are you in your answer?}
\end{quote}

They were provided with five potential answers, ranging from, \textit{``Very confident the samples are similar''} to \textit{``Very confident the samples are dissimilar.''}
The testing platform was implemented in JavaScript using D3~\cite{Bostock2011} and embedded into a Qualtrics survey accessed using the participant's web browser.

Our supplemental material includes full details of our survey instrument, including screenshots and complete wording.

\subsection{Dataset Generation}

We generated a collection of 1,800 pairs of data samples, 600 pairs per each sample size (see below). 
All dataset were drawn from a normal population.
The different pairs had the same number of items; 36, 144, or 1,000. 
This also creates distributions with some variation in noise level and occasional irregular features.
To generate interesting pairs of samples, where the samples in each pair are referred to as $A$ and $B$, we added a small constant to raise the mean for sample $B$ such that approximately 50\% of the pairs were significantly different at the $p = 0.05$ level using a t-test.
In addition to the actual data, we calculated the mean, standard deviation, and t-statistic for each of the pairs of samples.

Since our focus here was on the type of normally distributed data that are typical for parametric statistical testing, we only drew from the standard normal distributions when generating our datasets.
Furthermore, we opted not to manipulate the standard deviation for the samples in each pair; the standard deviation ranges from approximately 0.6 to 0.8 for both samples in a pair.
We leave an investigation of the impact of standard deviation on user performance in assessing significant differences for future work.

Note that the Student t-test statistic is associated with the t-distribution, which has a greater kurtosis (tailedness) than the Z (normal) distribution.
However, a fundamental assumption of the t-test is that the data samples being tested follow a more or less Gaussian (normal) distribution.

\definecolor{steelblue}{HTML}{4682B4}
\definecolor{sienna}{HTML}{A0522D}
\definecolor{colormix}{HTML}{a18983}

\subsection{Experimental Factors}

We modeled two factors in our experiment:

\begin{itemize}
    \item\textbf{Sample Size} ($S$): The number of items in the two samples being visualized.
    As the number of items increases, the samples will begin to approach the idealized distribution.
    We chose three levels: 36, 144, and 1,000 items.
    The first two levels represent typical dataset sizes that the general population may encounter in their daily life, whereas the third represents a large dataset where only small changes in the distribution will typically yield statistically significant differences.
    Also, since the 'statistical power' of a data sample is approximately equal to the square root of sample size, samples of 36 are about half the power of samples of 144, and 1/5th samples of 1,000. 
    
    \item\textbf{Visualization} ($V$): The visualization type used to represent the two data samples.
    Based on our review of the literature, we chose six distinctive visualization techniques (Figure~\ref{fig:teaser}):
    \begin{itemize}
        \item\textbf{Overlapping bell curves:} Two superimposed continuous filled-area charts visualizing fitted normal distributions of the underlying data samples (Figure~\ref{fig:stimulus-bell}).
        
        \item\textbf{Wilkinson dotplot:} Dotplots~\cite{Wilkinson1999dotplot} are unit visualization~\cite{Park2018b} versions of histograms where individual dots (circles) are stacked to represent each bin (Figure~\ref{fig:stimulus-dot}).
                
        \item\textbf{Bar histogram:} Two ``classic'' histogram where the aggregated number of data items for each bin is represented using a bar of uniform width, both drawn in the same visual space so that they overlapped (Figure~\ref{fig:stimulus-histogram}).
        
        \item\textbf{Stacked bar charts:} Two ``classic'' bar histograms as above, juxtaposed one over another with no overlap (stacked), and with each chart receiving half of the available vertical display space (Figure~\ref{fig:stimulus-stacked}).
        
        \item\textbf{Boxplot:} The box-and-whisker plot as pioneered by John W.\ Tukey~\cite{wickham201140} (Figure~\ref{fig:stimulus-box}).
        The central rectangle contains the middle half of the data (from the 25$^{th}$ to the 75$^{th}$ percentile), the median (50$^{th}$ percentile) is marked with a line, and the ``whiskers'' mark borders of wider percentiles, in this case the upper 10\% and lower 10\% of the data (the 10$^{th}$ and 90$^{th}$ percentiles).
        We did not visualize outlier data in our representation.
        
        \item\textbf{Strip plot:} A unit visualization~\cite{Park2018b} where each item is drawn as a short vertical line with opacity on the horizontal axes, i.e., with no vertical data encoding (Figure~\ref{fig:stimulus-strip}).
    \end{itemize}
    
    Since our task requires visualizing \textit{two} samples ($A$ and $B$) to allow comparisons, we opted to draw both overlapping bell curves, histograms, and dotplots in the same visual space using \colorbox{steelblue!50}{steelblue} and \colorbox{sienna!50}{sienna} colors at 50\% transparency.
    This gives rise to a special overlapping color (\colorbox{colormix}{brown} in Figures~\ref{fig:teaser} and \ref{fig:trial}) when visual marks for the samples overlap.
    For strip plots, we also used 50\% transparency, but overlapped the two plots only halfway, preventing overplotting (Figure~\ref{fig:stimulus-strip}).
    Finally, for boxplots, we separated the two plots entirely (Figure~\ref{fig:stimulus-box}).
    
\end{itemize}

The number of bins is a significant parameter for histograms.~\cite{journals/tvcg/CorrellLKS19}
We opted not to model this factor directly, instead keeping the extents of the horizontal axis constant (at $[-10, 10]$) and the number of bins constant (50).

\subsection{Experimental Design}

We used a mixed design, where each participant saw all data sizes but only one of the six available visualizations.
This allowed us to minimize the amount of training that would otherwise be required to instruct participants in the use of each visualization type.
The small total number of conditions enabled us to keep sessions shorter than 10 minutes in duration to minimize fatigue and maximize attention for crowdworkers.

Each trial pulled at random one of the pre-computed dataset pairs.
Trials for each dataset size were grouped to provide respondents the maximum opportunity to learn during the experiment.
They were presented with 10 trials of sample size 36 first, then 10 of 144, then 10 of 1000. 
Within each group, they were presented with two training trials, one which showed an example of a significant difference, the other not.   
It yielded the following design:

\bigskip

\noindent{\begin{tabular}{crl}
  & 3 & \textbf{Sample Size} $S$ (36, 144, 1000 samples)\\
  $\times$ & 1 & \textbf{Visualizations} $V$ (bell, stack, bar, dot, box, strip)\\
  $\times$ & 10 & repetitions + 2 training trials\\
  
  \hline

  & 30 & live trials per participant, plus 6 training trials\\

\end{tabular}}

\bigskip

For 500 participants, we planned to collect a total of 15,000 trials; instead, we ended up with 6,360 trials.
For each trial, we captured the correctness.
Correctness was defined as whether (1.0) or not (0.0) the participant correctly assessed a set of samples to be significantly different or not (based on a t-test at $p = 0.05$).
We also captured statistics of the actual datasets participants saw, such as the p-value of the corresponding t-statistic for the two samples.
A short set of optional text questions followed the trials, as well as a few demographic questions about respondents.   

\subsection{Procedure}

We recruited participants through Amazon Mechanical Turk.
Participants that fit the eligibility criteria opened the survey in a separate browser window.
At the end of their participation, they copied a unique completion code back into the Mechanical Turk interface, and were later paid as their work was checked. 
Participants were only allowed to conduct the experiment once.

Each session started with a consent form with waived signed consent. 
Failure to give consent terminated the experiment.
Participants were asked to confirm that they had no color vision deficiency.
Participants were allowed to abandon their session at any point in time.
Unfortunately, we were unable to pay participants who only completed a partial session.
We informed participants of this fact in the consent form before starting the session.

After each trial, the participant was given the correct answer after deciding whether or not a specific pair of samples were drawn from different populations or not.  
After trials, participants were asked demographic questions about their age, education level, and knowledge of statistics.

Each individual trial started with the display of the two samples and ended when participants clicked the one of the five answer buttons.
A progress bar at the top of the screen showed the study progress.
Amazon Mechanical Turk provides the option of rejecting specific respondents who failed to faithfully participate in a task, for example, by repeatedly clicking the same answer button to rush through the trials.
However, variability checks after collection showed no such cases in the Master Turker population we drew from.

Typical sessions lasted between 5 and 6 minutes in duration.
A few participants used significantly longer to complete their sessions, but our logs indicate that these participants took significant breaks between trials (presumably due to real-world interruptions).
We believe that the effective time spent on the experiment was no more than 10 minutes.

\subsection{Hypotheses}
\label{sec:hypotheses}

We preregistered the following hypotheses about our experiment:

\begin{itemize}
\item \textit{High-fidelity vs.\ low-fidelity visualization.} Participants will be more accurate at assessing significant differences ($p = 0.05$) between samples when using a high-fidelity visualization than when using a lower-fidelity visualization. 
    \begin{itemize}
    \item \textit{Bars and dots.} Bar histograms and Wilkinson dot plots have higher fidelity than all other representations tested, and will thus yield higher accuracy than all those other representations.
    \item \textit{Strip plots.} Strip plots are high-fidelity visualizations that will yield better accuracy than boxplots and idealized bell curves, but occlusion will result in lower accuracy than for bar histograms and dotplots, especially for large sample sizes.
    \item \textit{Boxplots.} Boxplots are an intermediate-fidelity visualization that will yield better accuracy than idealized bell curves.
\end{itemize}
\item\textit{Individual differences.} There will be non-uniformity in performance across individual participants; we anticipate that there will exist cohorts of participants with qualitatively different patterns of performance.
\end{itemize}

\section{Results}
\label{sec:results}

We analyzed the results primarily by computing correctness scores within analytic groups (based on Visualization Type, Sample Size of datasets, and p-value of the specific dataset pairs from each trial).
We judged a trial correct if its associated p value was $< .05$ and the respondent selected one of the two ``dissimilar'' categorical answers, or if the p value was $\ge .05$ and the respondent selected one of the two ``similar'' answers.
We then used bootstrapping~\cite{Efron1992} ($N = 1,000$ repetitions) on trials (aggregated average per participant) to compute 95\% confidence intervals.
We plotted these confidence intervals and used graphical inference to compare the different conditions.

\subsection{Overall Correctness Analysis}

As any given trial produces a dichotomous result, the 50\% cutoff is a useful comparison in the following to assess a visualization technique.
The nearer to 50\% correctness a technique produces in respondents, the nearer the results for that technique come to those we could expect through random guessing (and so the utility of that technique for statistical testing is diminished).

\begin{figure}[htb]
    \centering
    \resizebox{\linewidth}{!}{\includegraphics{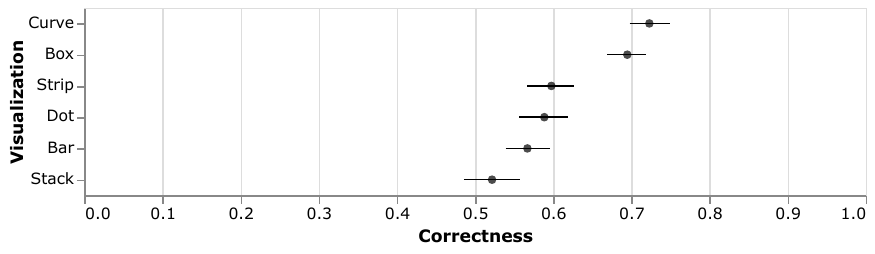}}
    \caption{\textbf{Correctness by visualization type.}
    Error bars show 95\% confidence intervals.
    }
    \label{fig:vis-correctness}
\end{figure}

Beginning with visualization type (Figure~\ref{fig:vis-correctness}), we find that respondents did appreciably better with the idealized bell curve plots and boxplots than with any other.
Respondents were correct about 70\% of the time with boxplots, and a bit more with idealized bell curves.
This not only fails to support our first hypothesis, it appears to directly contradict it.
These low-fidelity visualizations appear to allow respondents the best performance on this task, with the highest performance appearing in the lowest-fidelity visualization (the idealized bell curves).
Dot plots, bar histograms, and stacked bar histograms showed the worst performance, with stacked being essentially indistinguishable from chance.
Strip plots performed worse than normal curve and box plots, but at least as well as dot plots \& bar histograms, and better than stacked bar histograms.  

\begin{figure}[htb]
    \centering
    \resizebox{\linewidth}{!}{\includegraphics{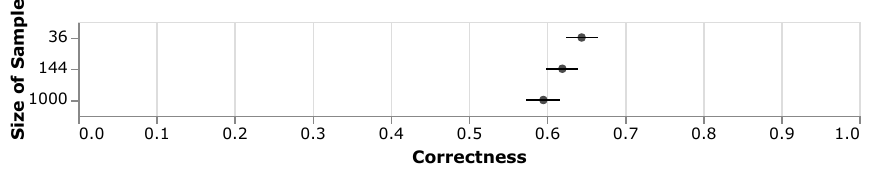}}
    \caption{\textbf{Correctness by sample size.}
    Error bars show 95\% confidence intervals.}
    \label{fig:size-correctness}
\end{figure}

We find that the size of datasets being compared had only a small impact on average correctness when considered as a stand-alone factor (Figure~\ref{fig:size-correctness}); certainly much smaller than the differences across visualization type.

\begin{figure}[htb]
    \centering
    \resizebox{\linewidth}{!}{\includegraphics{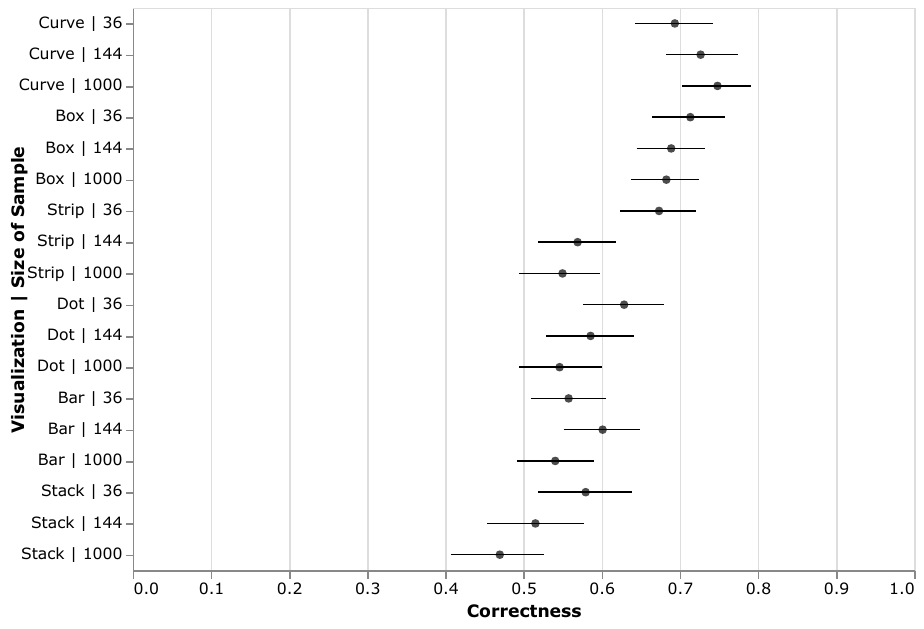}}
    \caption{\textbf{Correctness by visualization by sample size.} 
    Error bars show 95\% confidence intervals.}
    \label{fig:visxsize-correctness}
\end{figure}

There may be some degree of interaction between the size of compared datasets and specific visualization type (Figure~\ref{fig:visxsize-correctness}).
Normal bell curves may have a tendency to perform better at higher sample sizes.
In comparison, strip plots and dotplots did worse with higher dataset sizes; strip plots in particular appear to decline in effectiveness for this kind of comparison once the smallest sample size is exceeded.
This supports one part of our hypothesis regarding strip plots, namely, that at higher dataset sizes occlusion might become a problem.
However, dotplots may also suffer from occlusion.
The stacked bar histograms also appear to decline in effectiveness as sample sizes increase.  

\begin{figure}[htb]
    \centering
    \resizebox{\linewidth}{!}{\includegraphics{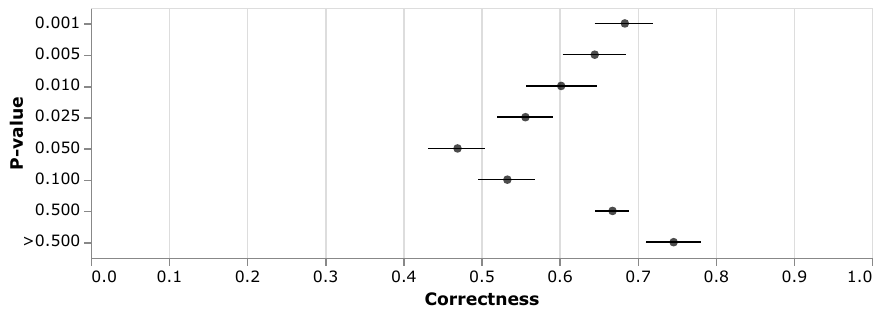}}
    \caption{\textbf{Correctness by p-value.} 
    Error bars show 95\% confidence intervals.
    The V-shape here suggests participant correctness decreasing as the p-value approaches 0.050.
    }
    \label{fig:p-correctness}
\end{figure}

We also examined the impact of the degree of difference between particular datasets respondents had to compare, expressed as p-values (Figure~\ref{fig:p-correctness}).
This should serve as a sort of measure of difficulty of the individual trial, where extremely low p-values will indicate dataset pairs with large mean differences, while p-values closer to 1 should indicate datasets with nearly total overlap. 

Indeed, there are large differences in correctness by p-value.
Respondents were far more likely to correctly identify datasets as dissimilar when the p-values were less than 0.001.
At the high end (p-values between .500 and 1), respondents were even more likely to make a correct choice (that datasets were ``similar'').
Trials with intermediate p-values, those between $p=.250$ and $p=.500$, gave respondents the most trouble; they did slightly worse than random chance, misidentifying dataset pairs in this p-value range as ``similar.''
Dataset pairs with p-values that approached significance ($0.050 < p < .010$) also presented a challenge, with respondents doing little better than chance on these.

\begin{figure*}[htb]
    \centering
    \subfloat[Bar histogram.]{
        \resizebox{0.3\linewidth}{!}{\includegraphics{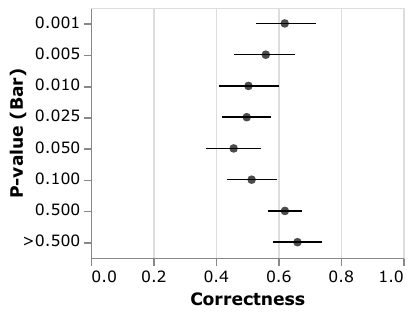}}
        \label{fig:barxp-correctness}
    }
    \subfloat[Wilkinson dot plot.]{
        \resizebox{0.3\linewidth}{!}{\includegraphics{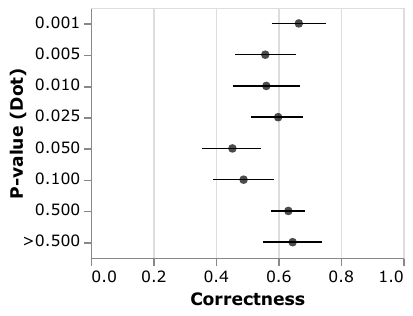}}
        \label{fig:dotxp-correctness}
    }
    \subfloat[Boxplot.]{
        \resizebox{0.3\linewidth}{!}{\includegraphics{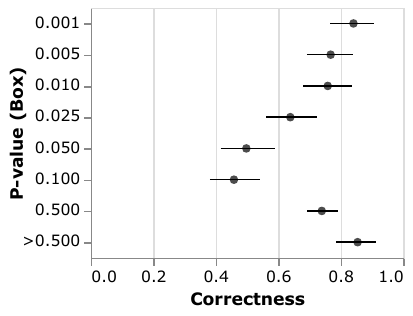}}
        \label{fig:boxp-correctness}
    }
    \\
    \subfloat[Strip plot.]{
        \resizebox{0.3\linewidth}{!}{\includegraphics{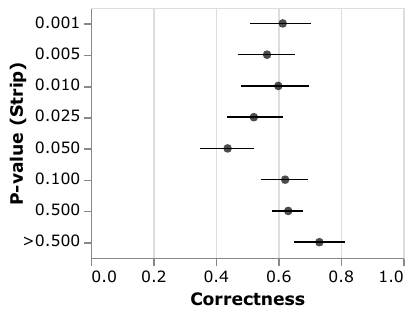}}
        \label{fig:stripxp-correctness}
    }
    \subfloat[Stacked bar histogram.]{
        \resizebox{0.3\linewidth}{!}{\includegraphics{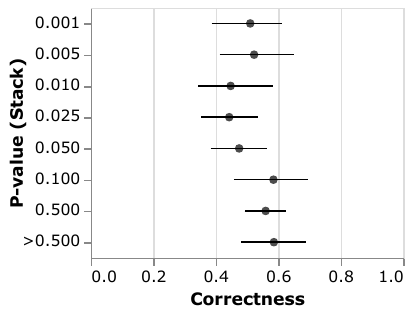}}
        \label{fig:stackxp-correctness}
    }
    \subfloat[Overlapping bell curves.]{
        \resizebox{0.3\linewidth}{!}{\includegraphics{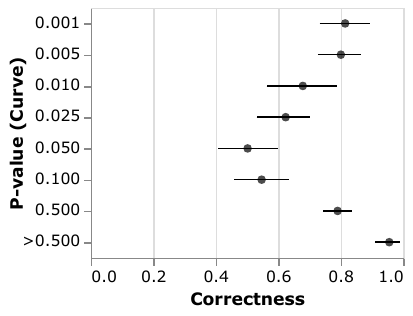}}
        \label{fig:curvexp-correctness}
    }
    \caption{\textbf{Correctness for visualization by p-value.}
    Error bars show 95\% confidence intervals.}
    \label{fig:visxp-correctness}
\end{figure*}

The interaction between visualization type and p-value confirms the general pattern of both individual factors, while also providing details that might be informative (Figures~\ref{fig:barxp-correctness}--\ref{fig:curvexp-correctness}).
The high average correctness respondents achieve with normal curve and boxplots comes from the very high percentage correctness at high and low p-values; respondents judged curves with p-values greater than .500 correctly more than 90\% of the time (Figure~\ref{fig:curvexp-correctness}).
Yet with intermediate values using these visualizations, respondents did essentially no better than chance guessing.
This in part reflects the nature of classic statistical tests, which in their simplest interpretation require a dichotomous response (significant or not), yet it also suggests a shortcoming of this approach, where even the most effective visualizations become ineffective when faced with difficult cases. 
Strip plots (Figure~\ref{fig:stripxp-correctness}), with which respondents performed at least as well or better than the different kinds of histograms, appear to get a boost from higher scores on the high p-value trials, but on other trials show quite modest scores.

\subsection{Individual Analysis}

\begin{figure}[htb]
    \centering
    \resizebox{\linewidth}{!}{\includegraphics{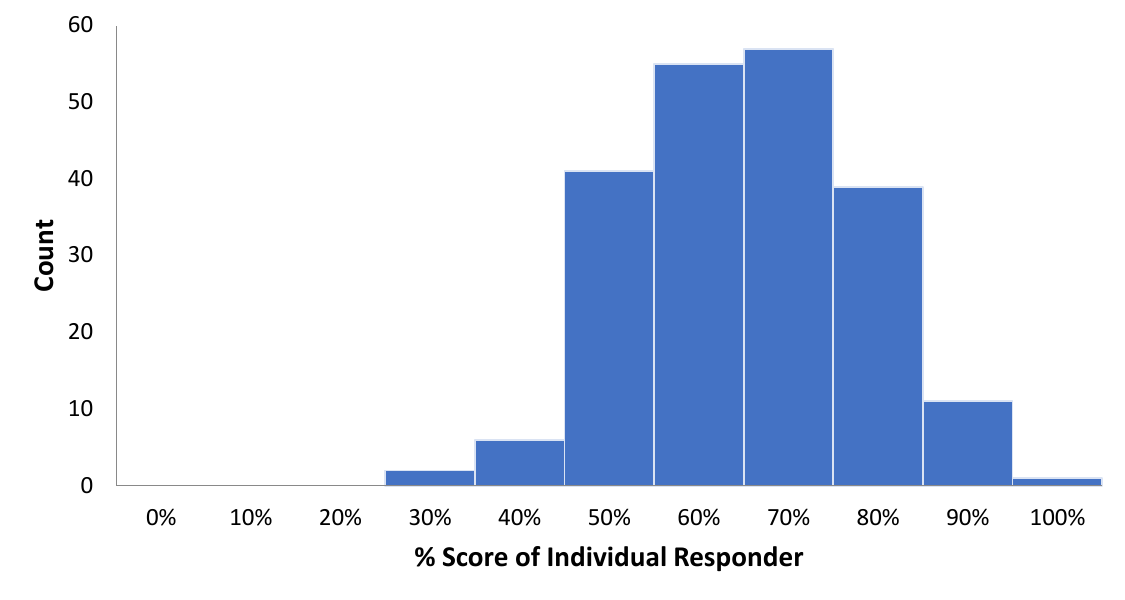}}
    \caption{\textbf{Distribution of individual respondent correctness.}
    Measured across all trials for all participants.
    }
    \label{fig:image_individual_performance}
\end{figure}

In our final hypothesis, we proposed that some group of individuals would show a propensity for higher performance on this task, that is, some people would have an ``eye'' for this kind of comparison.
However, our results do not support this contention (Figure~\ref{fig:image_individual_performance}).
The scores across participants appears to be normally distributed, with no clusters or modal humps that might suggest any structure other than that which random noise can explain.

It is possible that this normal distribution also represents a gradient in innate ability, which, like height or birth weight, has variation around some norm.
However, we would require repeated tests over time on the same respondents to confirm this interpretation.        

\subsection{Demographics and Participant Feedback}

Clusters in performance associated with demographic characteristics would also point to the possibility of consistent differences in performance among some group of individuals.
However, there is little evidence for such differences in these data (Table~\ref{tab:image_demo_tab}).

\begin{table}[htb]
    \centering
    \resizebox{\linewidth}{!}{\includegraphics{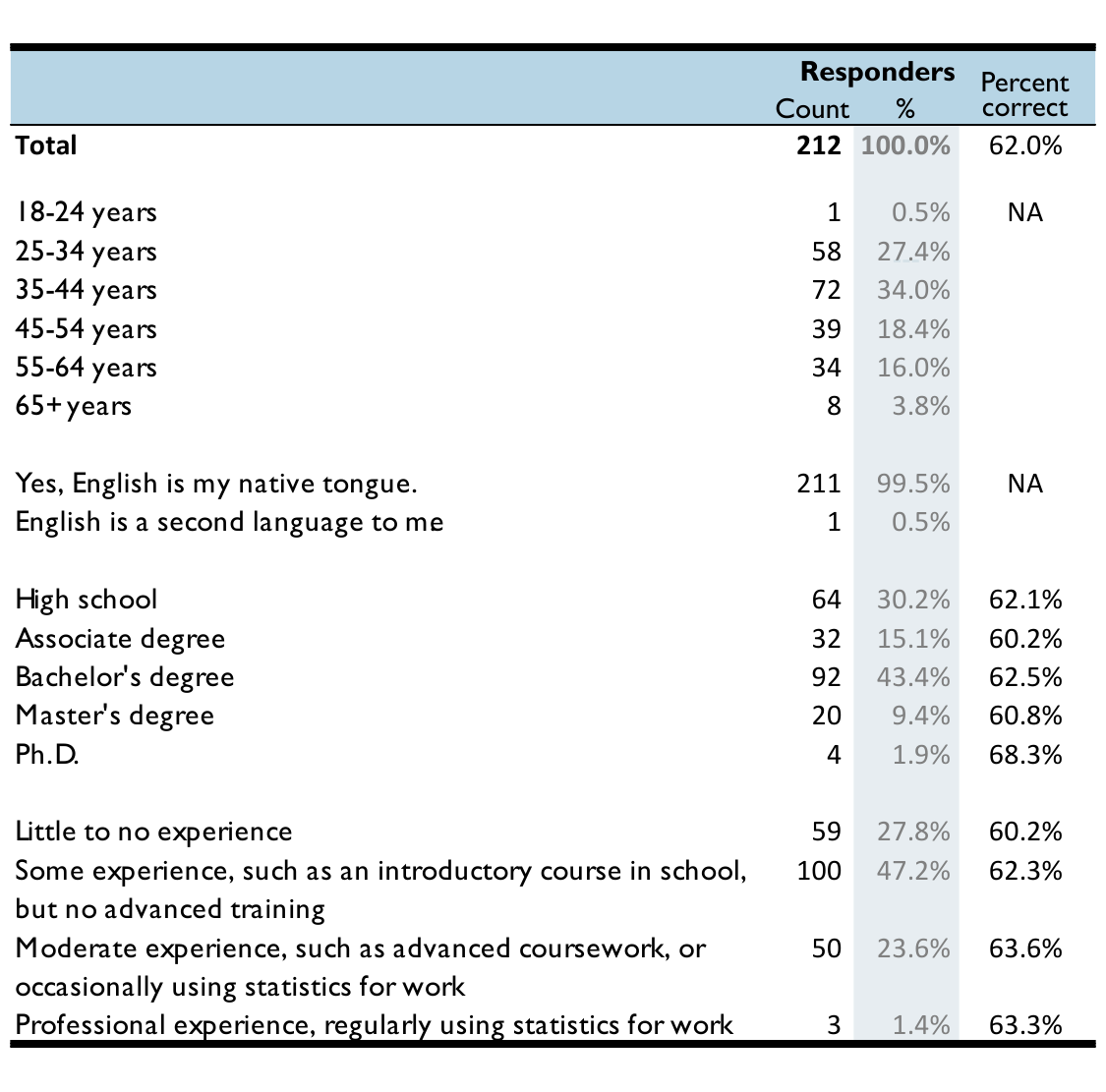}}
    \caption{\textbf{Participant demographics.}
    Collected from 212 participants in our crowdsourced experiment.
    }
    \label{tab:image_demo_tab}
\end{table}

Average percent correct achieved by respondents varied little by either education or prior experience with statistics (self reported). 

\subsection{Deviations from the Preregistration}

This experiment was preregistered in September 2020, but data collection only commenced in Summer 2021 due to what can only be expressed as pandemic fatigue.
We made the following deviations from the original study plans:

\begin{itemize}

\item\textbf{Added a visualization technique:} After feedback from colleagues, we added stacked bar histograms to the lineup of visualizations tested, bringing it up from 5 (as named in the preregistration) to 6.
The benefit of this change was to add a familiar and commonly encountered visual representation to the study, one which addressed the perhaps unfamiliar overlapping of bar histograms we employ here.
We do not anticipate that this had any ill effect on the validity or results of the experiment.

\item\textbf{Fewer participants:} We had originally aimed for 500 participants, with approximately 100 per visualization type. 
We ended up with only approximately 35 participants per visualization type because we raised our recruitment qualification to Turk Master Workers, which were both more expensive and more difficult to recruit.
However, we believe that the increased quality arising from these highly-rated workers made this deviation worthwhile.

\item\textbf{Increased compensation:} Recruiting Turk Master Workers meant a necessary increase in compensation from the \$1 listed in the preregistration. 
Again, this should have no detrimental effect on the experiment.

\end{itemize}

\section{Discussion}

We address our hypotheses as follows:

\begin{itemize}
    \item Overall, we find that what we call ``high-fidelity'' visualizations---bar histograms, dotplots, and strip plots---yielded lower accuracy for this task than ``lower-fidelity'' visualizations---idealized bell curves and boxplots.
    This is contrary to our hypothesis, where we postulated that the increased fidelity would yield better accuracy. \textit{(Rejected)}
    
    \begin{itemize}
        \item Bars (overlapping and stacked) as well as dotplots did \textbf{not} yield the highest accuracy; in fact, they arguably performed the \textbf{worst}. \textit{(Rejected)}
        
        \item Strip plots performed better than expected---certainly better than bar histograms and Wilkinsonian dotplots---but still yielded lower accuracy than bell curves and boxplots. \textit{(Rejected)}
        
        \item Boxplots, which we name ``intermediate-fidelity'' visualizations, did not yield better accuracy than idealized bell curves; there is little evidence for any difference in accuracy between the two techniques. \textit{(Rejected)}
        
    \end{itemize}
    
    \item We find non-uniformity in performance across individual participants; as our individual analysis showed, there are some participants who were able to complete this task much more accurately than others.  However, without additional rounds of data collection, we are unable to confirm that this variation reflects the innate or learned ability of particular individuals rather than some other source of random variation. \textit{(Inconclusive)}
    
\end{itemize}

We would argue that with so many of our original hypotheses rejected, these results are particularly interesting and worthy of further investigation.
Below we attempt to first explain and then generalize the findings.
Then we discuss what they mean for visualization design.

\subsection{Explaining the Results}

Our results contradict or fail to support our major hypotheses, which surprised us.
Rather than confirming the utility of detailed visualizations such as dotplots and strip plots, they suggest that aggregate visualizations are more appropriate for looking for differences between datasets.
This goes against our instincts as visualization researchers and practitioners; our bias is toward more detailed views rather than less.   
But should we have been so surprised?  

Calculating t-statistics for comparing two samples requires only the mean and standard deviation of each sample and the sample sizes.
All the information going into the calculation of either statistic is aggregate---just like the normal curves from our trials.
The curves are drawn by inputting a mean and standard deviation, and assuming a normal distribution.
T-tests also assume normal distributions in sampled populations.
Thus, in a very real sense, the normal curves presented respondents with the most direct visual analog to the t-test we used to judge their answers.

Boxplots are the next most aggregate visualization, and respondents using them performed nearly as well as those using normal curves.
However, boxplots provide respondents with no visual reference for estimating sample size, a vital consideration in statistical testing.
We attribute our respondents' success with this form at least in part to our clustering trials by sample size, and preceding each group of trials with worked examples to give them a ``feel'' for the critical degree of overlap.    
Respondents did worse with all the less aggregated and more detailed visualization forms that we tested.
We speculate that this additional detail may distract respondents from correctly identifying the critical degree of overlap, particularly for borderline cases.

The poor performance of our respondents when facing borderline cases, regardless of visualization type, may suggest another possibility.
We speculate that statistical significance, measured as a $p$-value of .05, may not align well with our intuitions for what constitutes a difference between two distributions.
This is also consistent with ample findings in the literature.~\cite{DBLP:journals/tvcg/HelskeHCYB21, DBLP:journals/tvcg/KaleKH21}
It presents a stumbling block to using visual methods for statistical testing where borderline cases are a possibility---typically not uncommon.

One source of explanation for these effects may be drawn from \textit{algebraic visualization design}.~\cite{DBLP:journals/tvcg/KindlmannS14}
For example, the overplotting in a strip plot for a large sample sizes is a confuser, violating the unambiguity principle from the algebraic visualization design framework.
Essentially, you can add any number of the same value to a sample with no change to the strip plot since the corresponding strips will all be drawn in the exact same position.
There is a similar effect at work for aggregating representations, such as bell curves, boxplots, and bar histograms: these visualizations do not convey the number of data points, so as long as the data distribution does not change, the visual representation will not change.
This again violates the unambiguity principle since, for example, increasing the number of items by an order of magnitude by drawing from the same population does not change the visualization.

Overall, our findings affirm results in graph comprehension,~\cite{Pinker1990, Shah2002} which notes that the inferences drawn from a chart are heavily dependent on the visual affordances of that chart.
For the specific case of the t-test, which is built upon normally distributed data, it is perhaps not surprising that visualizations that directly visualize this normal distribution---i.e., bell curves and boxplots, the so-called ``low-fidelity visualizations'' in our experiment---also yield the results most consistent with that test.

\subsection{Generalizing the Results}

As in any crowdsourced study conducted via online tools, the participant pool sets limits on the applicability of results to the broader population.
All our participants had internet access, a computer, and access to some form of electronic banking.
All participants were U.S.\ residents, and all but one spoke English as their native language.
However, participants in our study came from a broad range of age groups, education levels, and prior experience with statistics.
We believe these results may have modest general application.

We tested only some of the visual methods for displaying distributions.
However, the methods we tested vary in both degree of aggregation, detail, and visual complexity.
Due to this variability, we believe it is possible to use our results to gauge the likely performance of other visualization types.
For example, density plots are somewhere between bar histograms and normal curves in their degree of detail and visual complexity.
We speculate that their performance in these tasks would reflect this intermediate position. 

Our paper is focused on standard parametric t-tests, which are emblematic of traditional dichotomous testing. 
Of course, dichotomous testing is potentially harmful to science, and a more robust approach based on confidence intervals (CIs) and effect sizes is increasingly being suggested.~\cite{Cumming2005, Dragicevic2016}
However, even CI representations have been shown to easily lead to dichotomous thinking, which can be remedied with appropriate visual embellishment.~\cite{DBLP:journals/tvcg/HelskeHCYB21}
Nevertheless, given our focus in this work, we see such extensions as ideas for future work.

Finally, we believe the three dataset sizes we tested are representative of dataset sizes in a diversity of fields, from education, to opinion polls, to product acceptance testing.
However, these results may be unhelpful to those studying ``big data'' visualization problems, or other fields where data sizes are typically several orders of magnitude larger.
Similarly, while normally distributed data are common, many other distributions (e.g. bimodal or highly skewed), find uses in a multitude of fields.
Few of the results presented here may generalize to these.
Indeed, it may be that creating visual aids for inferential statistics that hinge upon more complex distributions requires the kinds of detailed representations that performed poorly in our experiment.

\subsection{Implications for Design}

Our study has direct implications for people tasked with displaying multiple overlapping data distributions, but also those seeking better distribution graphics in general.
Our central finding suggests that some tasks suffer from additional detail.
In this example, where the center and degree of spread around that center formed the primary basis for a decision, more detail impeded a good decision.
However, the responsible designer will have to weigh carefully whether they are working on such a problem, as the loss of detail prevents a user from making any other discoveries in the ``extraneous'' information.

Our results provide some insight into the visual affordances of the six different chart types studied in this paper.
While most of these affordances were previously known, they are worth restating here.
Generally speaking, high-fidelity visualizations yield a high amount of detail, but this can sometimes come at the detriment of overplotting for large datasets (or even modest ones), which in turn yields an inability to ``see the forest because of the trees.''
Low-fidelity visualizations, on the other hand, make use of data abstraction, which makes the general shape of the data---the ``forest''---discernible at the cost of the individual items---the ``trees.''
This means that while idealized bell curves and box plots make it straightforward to see the abstracted mean and extents of the data, bar histograms, Wilkinson dotplots, and strip plots are more useful for data that do not obey a standard Gaussian distribution.
For one thing, a high-fidelity visual representation will easily tell the viewer when the data is not normally distributed, a fact that an idealized bell curve or boxplot will obscure.
Having said all this, the design of our experiment is not really conducive to detailed insights about the affordances of the different visualizations.

While bell curves and boxplots performed best overall, one deficiency with both of these representations is that they do not directly show effect sizes or even the absolute size of the samples.
We have already noted that this is a violation of the unambiguity principle;~\cite{DBLP:journals/tvcg/KindlmannS14} in addition, such metrics are useful when utilizing these representations in practical tools.

Our findings also suggest that comparing distributions visually may be most reliable where differences are either large or very small.
Where differences are only moderate, judging the degree of difference by eye is challenging.
The visual designer might be able to meet that challenge with forms that magnify visual differences---a calculation process which would require further testing.
However, the fact that humans may have a low ability to judge the degree of separation where $p$-values are intermediate may itself have important implications for communicating statistical information to a general audience.

We also found that stacked bar histograms are inferior to overlapping bar histograms when considering two distributions despite the potential interference the overlap can cause. 
Furthermore, strip plots may be a good choice for displaying small datasets, but more aggregated visual forms (boxplots and bell curves) may be better when dataset size increases. 
We also find that boxplots remain a powerful tool for comparing distributions.
In other words, Tukey is still right.

\section{Conclusion and Future Work}

We have presented results from a crowdsourced evaluation investigating how well people can perform graphical inference of what essentially amounts to a t-test: comparing visual representations of two data samples to determine whether they are drawn from the same or different populations.
Contrary to our beliefs, we found that the more abstracted visual representations were more accurate for this purpose than the ones who showed an unaggregated version of the data: idealized bell curves and boxplots yielded better accuracy than histograms, dotplots, and strip plots. 
Furthermore, we found that these abstracted representations were unaffected (or even improved) by increasing sample sizes, something which was not true for the other representations.

However, upon further reflection, we note that this is perhaps not so surprising since the t-test we use as ground truth is based on idealized representations in the first place.
In other words, we view our main finding from this study that graphical formulations of statistical tests can be powerful, even to the point where they can stand in for traditional statistical tests for user populations that are not trained in inferential statistics, provided differences in samples are large enough. 
Overall, we view this as a victory for data visualization, but also a caution about the biases we hold that led to our original speculations that more detailed visualization techniques would prove superior.

We think that our study suggests a host of visual statistics work in the future: work that explores borderline cases, work that seeks to identify situations where detail helps and where it hurts, and work that explores which equation-based statistical methods may be effectively transformed into intuitive visual analogues.
Such methods will require more complex analogues that support greater degrees of embedded calculation.
In the future, we hope to coalesce all of these ideas into a general tool for inferential statistics aimed at laypeople, in essence enabling anyone to build on one of John Tukey's most favorite sayings:

\begin{quote}
\textit{``Far better an approximate answer to the \textbf{right} question, which is often vague, than the \textbf{exact} answer to the wrong question, which can always be made precise.''} --- John Tukey (1962)~\cite{Tukey1962}
\end{quote}



\bibliographystyle{plainnat}
\bibliography{overlapping-curves}

\end{document}